%% file: EXO_purity.tex
\documentclass[]{elsart}

\usepackage{graphicx}
\usepackage{rotating}
\usepackage{supertabular}
\usepackage{longtable}
\usepackage{lscape}
\usepackage{epsfig}
\usepackage{rotating}
\usepackage{ifthen}
\usepackage{multicol}

\usepackage{amssymb}
\usepackage{xspace}

 \usepackage{lineno}

\newcommand{\xe}{\ensuremath{\rm ^{136}Xe}\xspace}
\newcommand{\K}[1]{\ensuremath{\rm ^{#1}K}\xspace}
\newcommand{\U}[1]{\ensuremath{\rm ^{#1}U}\xspace}
\newcommand{\Th}[1]{\ensuremath{\rm ^{#1}Th}\xspace}
\newcommand{\Pb}[1]{\ensuremath{\rm ^{#1}Pb}\xspace}
\newcommand{\Po}[1]{\ensuremath{\rm ^{#1}Po}\xspace}
\newcommand{\Tl}[1]{\ensuremath{\rm ^{#1}Tl}\xspace}
\newcommand{\Ac}[1]{\ensuremath{\rm ^{#1}Ac}\xspace}
\newcommand{\Bi}[1]{\ensuremath{\rm ^{#1}Bi}\xspace}
\newcommand{\Cs}[1]{\ensuremath{\rm ^{#1}Cs}\xspace}
\newcommand{\KTHU}{K, Th, and U\xspace}
\newcommand{\hnot}{\ensuremath{\rm HNO_3}\xspace}
\newcommand{\Ar}[1]{\ensuremath{\rm ^{#1}Ar}\xspace}
\newcommand{\degree}{\ensuremath{^{\circ}}\xspace}
\newcommand{\degrees}{\degree}

\newcommand{\figwidth}{\columnwidth}
\newcommand{\ife}[3]{\ifthenelse{\equal{#1}{#2}}{#3}{{}}}
\input{linemap}

\begin{document}

\begin{frontmatter}

\title{Systematic study of trace radioactive impurities in candidate construction materials for EXO-200}

\author[bama]{D.S.~Leonard},
\author[INMS]{P. Grinberg},
\author[UN]{P.~Weber\thanksref{PW}},
\author[UN]{E.~Baussan\thanksref{EB}},
\author[bama]{Z. Djurcic\thanksref{ZD}},
\author[bama]{G.~Keefer},
\author[bama]{A.~Piepke},
\author[SU]{A.~Pocar},
\author[UN]{J.-L.~Vuilleumier},
\author[UN]{J.-M.~Vuilleumier},
\author[ITEP]{D.~Akimov},
\author[CU]{A.~Bellerive},
\author[CU]{M.~Bowcock},
\author[SLAC]{M.~Breidenbach},
\author[ITEP]{A.~Burenkov},
\author[SLAC]{R.~Conley},
\author[SLAC]{W.~Craddock},
\author[ITEP]{M.~Danilov},
\author[SU]{R.~DeVoe},
\author[CU]{M.~Dixit},
\author[ITEP]{A.~Dolgolenko},
\author[CU]{I.~Ekchtout},
\author[CSU]{W.~Fairbank~Jr.},
\author[LU]{J.~Farine},
\author[SU]{P.~Fierlinger},
\author[SU]{B.~Flatt},
\author[SU]{G.~Gratta},
\author[SU]{M.~Green},
\author[UM]{C.~Hall},
\author[CSU]{K.~Hall},
\author[LU]{D.~Hallman},
\author[CU]{C.~Hargrove},
\author[SLAC]{R.~Herbst},
\author[SLAC]{J.~Hodgson},
\author[CSU]{S.~ Jeng},
\author[SU]{S. Kolkowitz},
\author[ITEP]{A.~Kovalenko},
\author[ITEP]{D.~Kovalenko},
\author[SU]{F.~LePort},
\author[SLAC]{D.~Mackay},
\author[UCI]{M.~Moe},
\author[SU]{M.~Montero~D\'{i}ez},
\author[SU]{R.~Neilson},
\author[SLAC]{A.~Odian},
\author[SU]{K.~O'Sullivan},
\author[UN]{L.~Ounalli},
\author[SLAC]{C.Y.~Prescott},
\author[SLAC]{P.C.~Rowson},
\author[UN]{D.~Schenker},
\author[CU]{D.~Sinclair},
\author[SLAC]{K.~Skarpaas},
\author[ITEP]{G.~Smirnov},
\author[ITEP]{V.~Stekhanov},
\author[CU]{V.~Strickland},
\author[LU]{C.~Virtue},
\author[SLAC]{K.~Wamba},
\author[SU]{J.~Wodin\thanksref{JW}}

\address[bama]{Department of Physics and Astronomy, University of Alabama, Tuscaloosa AL, USA}
\address[INMS]{Institute for National Measurements Standards, National Research Council Canada, Ottawa ON, Canada}
\address[UN]{Institut de Physique, Universit\'{e} de Neuchatel, Neuchatel, Switzerland}
\address[SU]{Physics Department, Stanford University, Stanford CA, USA}
\address[ITEP]{ITEP, Moscow, Russia}
\address[CU]{Physics Department, Carleton University, Ottawa ON, Canada}
\address[SLAC]{Stanford Linear Accelerator Center, Menlo Park CA, USA}

\address[CSU]{Physics Department, Colorado State University, Fort Collins CO, USA}
\address[LU]{Department of Physics, Laurentian University, Sudbury ON, Canada}
\address[UM]{Department of Physics, University of Maryland, College Park MD, USA}
\address[UCI]{Physics Department, University of California, Irvine CA, USA}

\thanks[PW]{Now at Center for Ion Beam Physics, Swiss Universities for Applied Physics, La Chaux-de-Fonds, Switzerland}
\thanks[EB]{Now at IPHC, Universit\'e Louis Pasteur, CNRS/IN2P3, Strasbourg, France}
\thanks[ZD]{Now at Columbia University, NewYork NY, USA}
\thanks[JW]{Now at Stanford Linear Accelerator Center, Menlo Park CA, USA}

\begin{abstract}

The Enriched Xenon Observatory (EXO) will search for double beta decays of \xe.  
We report the results of a systematic study of trace concentrations of radioactive impurities in a wide range of raw materials and finished parts considered for use in the construction of EXO-200, the first stage of the EXO experimental program.  Analysis techniques employed, and described here, include direct gamma counting, alpha counting, neutron activation analysis, and high-sensitivity mass spectrometry.


\end{abstract}

\end{frontmatter}


\section{Introduction}
\label{sec:Intro}
\begin{figure}[htb]\centering
\includegraphics[width=\figwidth]{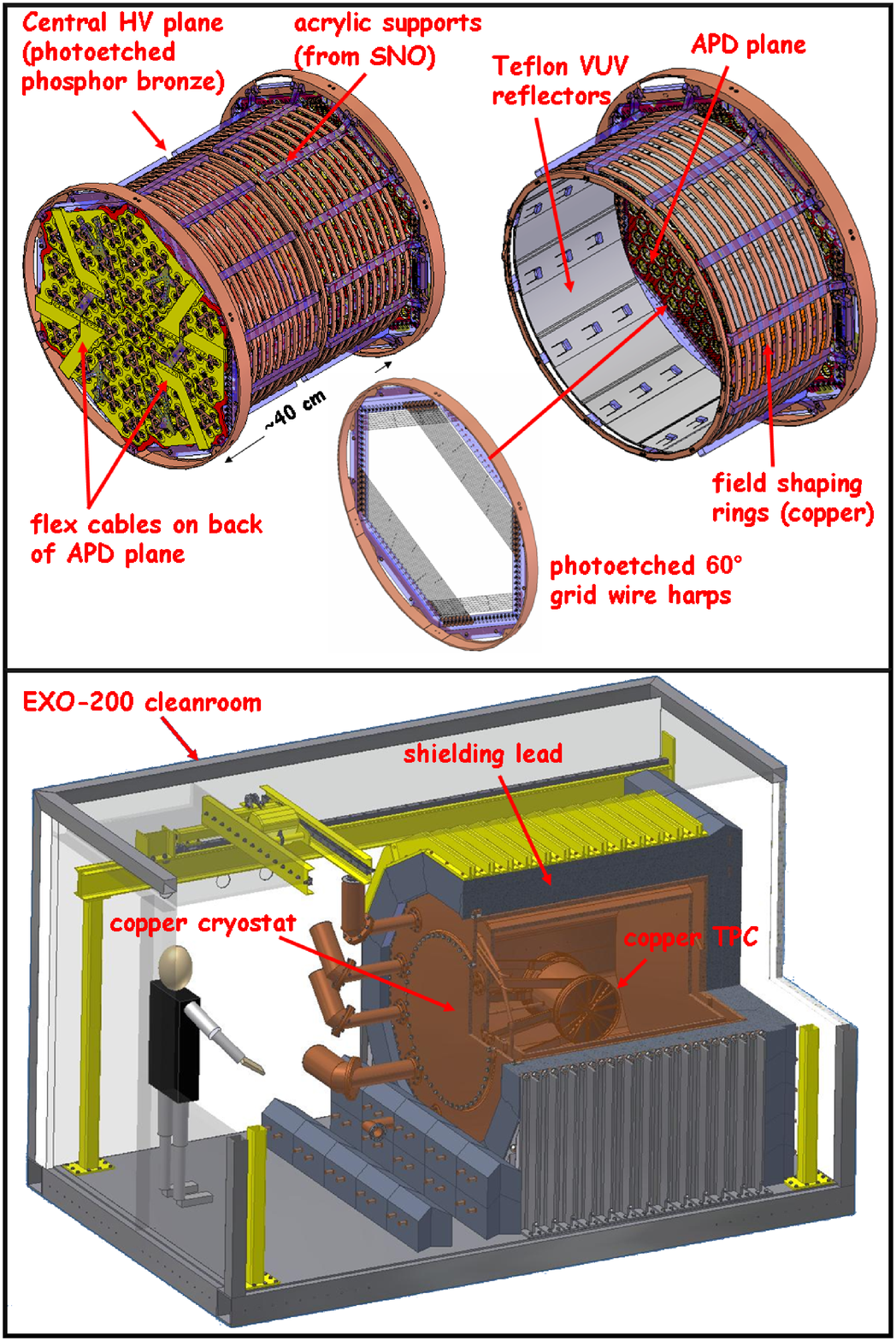}
\caption{Top panel: The EXO-200 TPC with major internal parts shown. Bottom panel:  Cut-away view of the TPC housed in the shielded cryostat.   The cleanroom is underground in the Waste Isolation Pilot Plant (WIPP) near Carlsbad, NM. with an overburden of  $\rm 1585^{+
11}_{-6}~m.w.e.$~\cite{esch}. }
\label{fig:EXO200}
\end{figure}

This work was motivated by the Enriched Xenon Observatory (EXO), a multi-stage experimental research program with the purpose of detecting rare double beta decays in \xe~\cite{EXO}.  With EXO-200, the first stage of the project, we will search for these decays in an underground cryogenic time-projection chamber (TPC) filled with approximately 200~kg of liquid xenon enriched to 80\% in \xe.  The EXO-200 detector is shown in Fig.~\ref{fig:EXO200} and described in more detail in Ref.~\cite{teflon}.  To reach the desired half-life sensitivity of about $\rm 6\times10^{25}~yr$ for the 0-neutrino decay mode,  the background rate of candidate events (single-site events within 2-$\rm\sigma$ of the Q value of 2458~keV~\cite{XeQ})  cannot exceed about 20 events per year.  The background rate for the 2-neutrino decay mode should not exceed about $30\,000$ events per year between a threshold of approximately 400~keV and the Q value.  Similarly strict background requirements are common to other double beta decay and rare-event detectors.  To achieve very low background rates, all materials present in the detector, and even some external materials that are in the path of the xenon criculation or the cryogenic fluid handling, must be selected to have very low intrinsic concentrations of radioactive impurities, especially the naturally occurring elements \KTHU.   We performed an extensive campaign of radiopurity measurements to search for and certify a library of suitable materials for every component of the EXO-200 detector.  Special attention was given to materials considered for massive parts of the detector and for materials near the active volume or in contact with the xenon.

Our techniques for measuring intrinsic purities include mass spectrometry (MS), neutron activation analysis (NAA), and direct gamma counting.  In addition,  alpha counting was used to determine \Pb{210} concentrations in lead.   Measurement techniques used for specific samples were selected based on a number of factors and constraints.  Both NAA and MS can be used with small sample masses, on the order of a gram. In comparison to NAA, mass spectrometry is less expensive and faster.  Glow-discharge mass-spectrometry (GD-MS) gives results for many elements simultaneously but has relatively poor sensitivities and can only be used with conductive or semi-conductive solids.  In contrast, inductively-coupled plasma mass spectrometry (ICP-MS) can have much higher sensitivities, but the difference is large only if preconcentration procedures are used.  Only materials which are chemically compatible with these procedures are candidates for ICP-MS.  For our U and Th ICP-MS analyses, this implies that samples must be soluble in \hnot or possibly in some other acids.  NAA can be performed on small samples of many types of materials and can, in some cases, produce the best sensitivities of all known techniques, but the process is relatively costly and slow.  Depending on the composition of the material, activation signals from the primary constituents or from uninteresting trace contaminants can interfere with the signals of interest, often resulting in higher than ideal or even useless sensitivity levels.  Finally, direct gamma counting can be used to simultaneously measure levels of \KTHU (as well other radioactive isotopes) in virtually any material, but requires large sample masses and long measurement times. 

MS and NAA assay directly the \Th{232} and \U{238} concentrations.  However, these isotopes decay initially through low-energy decays below the EXO-200 threshold and through alpha emissions which, as well has having very short ranges, can be rejected by comparison of ionization and scintillation signals.    It is subsequent decays of particular daughter isotopes which produce high energy gamma emissions that can penetrate into the detector's interior and contribute to the EXO-200 backgrounds.   If all isotopes in a decay chain remain within the same volume, then secular equilibrium is achieved.  The activity, and thus concentration, of any isotope can then be inferred from a measurement of the activity or concentration of any other isotope within the same decay chain.  However, chemical processes or outgassing can remove radioactive daughter isotopes from materials, thus breaking secular equilibrium. Gamma counting has the advantage that it measures the gamma intensities of the decay products relevant to the EXO-200 backgrounds. For consistency though, data from all techniques used are reported here as total concentrations of the element corresponding to the progenitor isotope.  Gamma counting results have thus been converted assuming secular equilibrium in the decay chain.  For each of \KTHU, natural terrestrial abundance ratios were used to covert from isotopic to total elemental abundances.

Differences in the measurement techniques were weighed against our resources and our requirements for each material.  Results are listed in Tables~\ref{tab:Lead} and~\ref{TheTable}.  The tables also include some data (clearly identified) produced by commercial laboratories using similar techniques.

The liquid-xenon TPC of the EXO-200 detector is contained in a thin-walled copper pressure vessel.   However, in various phases of the design, we have considered other materials for construction of the vessel.  These included quartz, high-purity copper,  fluoropolymers, and other plastics.  Using NAA, described in Section~\ref{sec:NAA}, we found some fluoropolymers, (especially DuPont Teflon TE-6472) in their raw forms, to be among the cleanest solids that we are aware of, both from a radio-purity perspective and also from a general chemical-purity perspective.  After careful review and oversight of the sintering process, we also found that finished parts could be produced cleanly.  Table~\ref{TheTable} entries~\idx{185}--\idx{204} list results for various fluoropolymers in raw form as well as finished products, in some cases both before and after reviewing the sintering process.  Detailed descriptions of functional studies of fluoropolymer test vessels are described in Ref.~\cite{teflon}. 

The TPC contains wire grids for charge collection, and avalanche photo diodes (APDs) for detection of scintillation light. Copper-clad flexible circuits, used in the manufacturing of various signal cables, were analyzed in two steps.  First the copper was dissolved in \hnot and measured using ICP-MS.  The remaining polyimide substrate was then measured using NAA.  Germanium counting was also used to analyze whole, unaltered samples of these circuits. Commercial photo-etching processes were investigated for production of the grid wires and APD contacts.  Again, by careful oversight, including  testing and selection  of chemicals (see Table~\ref{TheTable} entries~\idx{26}--\idx{32}) used in the etching process, we were able to improve the cleanliness of the products.   Materials used in the construction of the APDs, including the silicon substrate and the evaporated metal coatings, were individually studied to improve on the cleanliness of the product.  Results for APDs produced by Advanced Photonix Inc. (API), using their default material suppliers, are shown in Table~\ref{TheTable} entries \idx{63}--\idx{72}.  Entries \idx{67}--\idx{69} show results for aluminum used in the APDs (used in both the Au contacted and Al contacted APDs) and supplied by API's default supplier.  Entries \idx{53}--\idx{58} list measurement results for aluminum replacement candidates procured by EXO.

The EXO-200 TPC is surrounded by a cryogenic heat transfer fluid (HTF)  which serves as a thermal bath and also as a hermetic clean radiation shield. The HTF is contained and cooled in a large copper cryostat surrounded by lead shielding and placed underground in the Waste Isolation Pilot Plant (WIPP) near Carlsbad, NM.  Because of the large mass of the HTF (approximately 4.2 tonnes), we made extra efforts to select and certify candidate materials for this use.  Measurements of the final candidate material, 3M HFE-7000, are described in more detail in Section~\ref{sec:HFE}.  Finally the copper cryostat containing the HTF is surrounded by a thick layer of lead shielding.  Measurement and selection of lead at every stage of the production process is described in Section~\ref{sec:Lead}.

Many other raw  materials and parts were studied including vacuum grease, paint, nuts, bolts, and o-rings.  All measurements were selected and prioritized as needed for the specific construction and scheduling needs of EXO.  In some cases,  materials with unknown origins were tested for use in EXO-200.  We report these results here because they give useful impressions of the general trends and achievable purity levels in various types of materials.

\section{Handling and Surface Cleanliness}
\label{sec:Surface}

Surface cleanliness of all samples is a concern for all analysis methods used.  Although we normalize most results to the sample masses, assuming that the concentrations are intrinsic to and homogenous throughout the materials, it is clear that positive readings can also come from surface contaminants.  

For  measurements of raw materials, extensive surface cleaning was performed and samples were handled in cleanroom environments.  Typical cleaning procedures included extended soaking in solvents with sonic agitation, as well as extended soaking in 0.1 to 3~M (or higher for NAA preparation of some materials) nitric acid solutions with certified purity levels.  Within a few variations, depending on the material properties, cleaning procedures were followed consistently to assure reproducibility.  The effectiveness of the cleaning procedures is demonstrated by the stringent contamination limits set by the most sensitive measurements of the cleanest materials.

When analyzing finished parts (nuts, bolts, o-rings, etc.), care was taken to match the cleaning procedures  of the analysis techniques to those that were or will be used for the actual installed parts.  In most cases these procedures were similar to those used for analysis of raw materials.  In some cases however, due to time, volume, or integrity constraints, it was not possible to perform the full cleaning procedure.  Especially for non-critical parts with relatively high tolerable activities, if the parts could be well certified using less stringent cleaning procedures, then some burden or concern could be removed from the construction process. 

Specifically, most EXO-200 finished parts  were surface cleaned using the following treatment with high purity solvents targeting organic, metallic, and ionic surface impurities:

\begin{enumerate}
\item Acetone rinse followed by ethanol (or methanol) rinse to remove any grease
left over from machining or touching. Small parts underwent ultrasonic cleaning
for about 15~minutes while immersed in each solvent, often followed each time by a rinse in the same solvent.
\item 0.1 to 1~M \hnot rinse to dissolve metallic contamination.
For metallic parts light etching results in removal of the 
surface.  Small parts were fully immersed in acid.
\item Deionized water rinse following the etch. Small parts were dried in a vacuum oven;
large parts were left to dry in a cleanroom. 
\end{enumerate}

For thin photo-etched phosphor-bronze TPC parts, we found high levels of U which appeared to be surface contamination resulting from the commercial etching procedure.  We were able to reduce the U levels by requesting the use of fresh chemicals in the etching process (compare  entries~\idx{29} and~\idx{30} of Table~\ref{TheTable}) in combination with use of an expanded cleaning procedure which we developed  to maximize cleaning effectiveness while minimizing loss of material.  These parts were first rinsed in methanol followed by deionized water and then rinsed three times in 3~M~\hnot for 10 minutes, followed each time by a further rinse in deionized water.  This procedure was used to obtain the result in Table~\ref{TheTable} entry \idx{24}; a 15\% loss of sample mass was observed.

After surface cleaning, parts were double bagged (when practical according to size) and were only touched with disposable powder free gloves to prevent re-contamination.  

\begin{table}[bt]
\begin{tabular}{l l l}
Sample                                           & Th [pg/cm$^2$] & U [pg/cm$^2$]  \\ \hline
Xenon piping, burst disk, after cleaning$^a$.                      & $5.8\pm 0.2$   & $1.2\pm 0.1$   \\
Bottom inner surface of outer cryostat, before cleaning$^a$. & $380\pm 5$     & $140\pm 3$     \\
Top inner surface of outer cryostat, before cleaning$^b$.    & $2.6\pm 0.2$   & $0.46\pm 0.04$  \\
Top outer surface of outer cryostat, after cleaning$^b$. & $0.60\pm 0.09$ & $0.21\pm 0.05$ \\
Jehier super insulation from top of cryostat, after alcohol rinse$^b$.                   & $5.9\pm 0.2$   & $1.6\pm0.1$\\
Wessington Cryogenics HFE storage dewar,\tabularnewline manufacturer cleaning with wipes soaks and pressure washing$^a$.                                & $215\pm 12$    & $43\pm 6$      \\
\hline
\end{tabular}
\vskip 5mm
\caption{Th and U surface contamination
as determined by wipe testing. Blanks have been subtracted.\newline
$^a$ The filter was wetted with ethanol before wiping.\newline
$^b$ The filter was wetted with 0.1~M \hnot before wiping.
}
\label{tab:wipe_res}
\end{table}

Surface cleanliness of large parts was directly verified by means of wipe testing with 
Whatman 42 paper filters.  The Th and U content of the filters
was determined, after ashing, by means of ICP-MS.
To reduce the blank concentrations, the filters were
soaked for 24~h in 1~M \hnot followed by thorough rinsing
with deionized water and drying. Analysis of blank filters 
yielded Th and U masses of $0.27\pm 0.06 $ and $3.0\pm 0.6$ ng 
per filter respectively before acid cleaning and yielded $0.31\pm 0.02$ and $0.15\pm 0.03$ ng after acid cleaning.
To facilitate transfer of surface activity onto the filter paper,
the surface to be tested was wetted with either alcohol or
0.1~M nitric acid. The fluid was spread over some surface area
and then absorbed with the filter paper. 
Blank filters were always analyzed to control the filter paper background.

After drying, the filter paper samples were  accurately weighed, transferred into porcelain crucibles, and placed in a muffle furnace at 200\degrees C for 20~minutes. The temperature of the furnace was then gradually raised at a rate of approximately  10\degrees C/min with pauses for 30 minutes at 300\degrees C and 400\degrees C and finally raising the temperature to 500\degrees C where it was maintained for about one hour.  The crucibles were then removed from the furnace and allowed to cool down. After the samples were cooled, 10~ml of 3~M nitric acid was added to each crucible, evaporated to near dryness on a hotplate, and reconstituted to 10 ml with 0.5\% nitric acid. The Th and U content of the filters was determined by means of ICP-MS. Recovery tests were performed where the filter papers were spiked with known concentrations of the analytes.  Complete recovery 
($\rm>$ 94\%) was obtained.

Table~\ref{tab:wipe_res} summarizes the results for
various wipe tests performed on EXO-200 samples.

\section{Mass Spectrometry}
\label{sec:MS}

Inductively-coupled plasma mass spectrometry  offers sub pg/ml detection limits for U and Th with minimal analysis time. However, one of the main limitations of this technique is the need for sample preparation prior to analysis, as higher levels of matrix components can give rise to deposition of matrix constituents on the sampler and skimmer cones of the spectrometer. Thus, a dissolved sample may need to be diluted in order to lower its total dissolved solids content to $\rm<1\%$, clearly degrading achievable detection limits. One way of overcoming this drawback is to undertake prior separation of the analytes from the matrix. Several methods have been used for this purpose including coprecipitation~\cite{Cho00,Ero98,Loz99}, liquid-liquid extraction~\cite{Bec98,Bec99,Tag95}, distillation~\cite{Vij92}, and ion-exchange~\cite{Car99,Cha03} which was used in this work (see~Sec.~\ref{sec:ICP-MS}). These techniques often require longer analysis times and give rise to additional analytical problems, including contamination during sample pretreatment and increased blank levels, which must be carefully controlled. 

Direct analysis of solid samples can be advantageously performed, without chemical sample preparation, by glow-discharge mass spectrometry (GD-MS) \cite{Myk90,Win04} wherein the sample functions as the cathode of the discharge, making this approach particularly suitable for the analysis of high-purity metals.

\subsection{ICP-MS}
\label{sec:ICP-MS}
An ELAN DRC II (dynamic reaction cell) ICP-MS (Perkin-Elmer Sciex) equipped with a cyclonic glass spray chamber and a pneumatic nebulizer (Meinhard) was used for the determination of U and Th with sensitivities as low as $10^{-12}~g/g$. The ICP-MS operating parameters were selected to maximize sensitivity for U and Th.
 Samples were acid digested and then the analytes separated using a chromatography resin (UTEVA from Eichron). The U fraction was eluted from the resin with 15 ml of 0.02 M HCl, and the Th fraction was subsequently eluted with 30~ml of 0.5~M oxalic acid. Both fractions were evaporated to near dryness; the thorium fraction was decomposed with 24~ml of a mixture of concentrated \hnot/30\% and $\rm H_2O_2$ (1:1). Both fractions were reconstituted to 2 ml with 0.5\% nitric acid. U and Th were measured at masses 238 and 232 respectively. The procedure is described in more detail in Ref.~\cite{Gri05}.

For the determination of K concentrations at levels below $\rm10^{-6}$~g/g, samples were acid digested, diluted 100-fold and subsequently analyzed by DRC-ICP-MS using an ultrasonic nebulizer as a sample introduction system. \K{39} suffers from interferences from $\rm ^{38}Ar\,^1H^+$ and \Ar{40}. These interferences were overcome with the use of chemical resolution, a process used to selectively remove interfering polyatomic or isobaric species from the ICP-MS ion beam using controlled ion-molecule chemistry. Ammonia (at $\rm0.5~ml/min$ flow rate) was used as a reaction gas as it reacts with the Ar-based interferences to form new species that do not interfere with \K{39}. Samples were introduced to the ICP-MS with an ultrasonic nebulizer in order to further decrease the hydride formation and consequently the interference to \K{39}. The samples were nebulized by a piezoelectric crystal transducer. The nebulized aerosol was passed through a heated chamber and condenser where the solvent vapor was removed.

Uncertainties in the ICP-MS results were dominated by variability in the subtracted backgrounds, as observed by analysis of clean digestion acids, and to a lesser extent by scaling errors due to calibration uncertainties.  All known uncertainties are included in the tabulated results.

\subsection{GD-MS}
A VG 9000 glow discharge mass spectrometer (Thermo Electron Corp., UK) was used for direct analysis of solid samples (conducting and semiconducting solids). The instrument is capable of detecting impurities directly in the solid from the percent level down to below $\rm 10^{-9}~g/g$ in a single run, allowing for a rapid turn around of submitted samples. It relies on a DC glow discharge ion source coupled to a high resolution magnetic sector analyzer in reverse Nier Johnson geometry with electron multiplier detection at nominal resolution $R=3000$, where $R\equiv \rm{m}/\delta \rm{m}$, and $\delta \rm{m}$ is the resolvable mass difference at an ion mass of $\rm m$. Atoms were sputtered in a low-pressure DC argon discharge, subsequently ionized in this plasma, and extracted into the mass spectrometer for separation and detection. Test portions of the samples were prepared by cutting pins of approximately $ \rm 2.5  \times 2.5\times 20~\rm{mm}$ and subjecting them to a careful surface leach in dilute ultrapure nitric acid. Following a rinse with ultrapure water, the samples were permitted to air dry in a laminar-flow class-100 clean-bench, where they were subsequently mounted into the DC ion source. Once under vacuum, and then Ar purged, the glow discharge was ignited. Any surface contamination on the test samples was removed by a 30 minute pre-burn in the plasma before data acquisition was initiated. Typically 300 mass spectral scans were acquired, each having a 50~ms dwell or integration time. The GD-MS instrument was calibrated with the use of a variety of reference materials used to establish relative sensitivity factors to provide semiquantitative analysis, results of which are deemed to be (conservatively) within a factor of 2 of the real value of the concentration of the analyte. GD-MS is free from the matrix dependence response plaguing most other elemental analysis techniques, minimizing the need for matrix matched standards.

\section{Gamma Counting}
\label{sec:gamma}

For EXO, gamma counting was used primarily for small non-critical components, such as screws, washers, etc., which contribute little to the EXO-200 detector mass.  This technique was also used for cross checking other measurements of bulk materials,  in particular measurements of the copper for the TPC vessel and cryostat, and of the shielding lead.  

\begin{figure}[htb]\centering
\includegraphics[width=\figwidth]{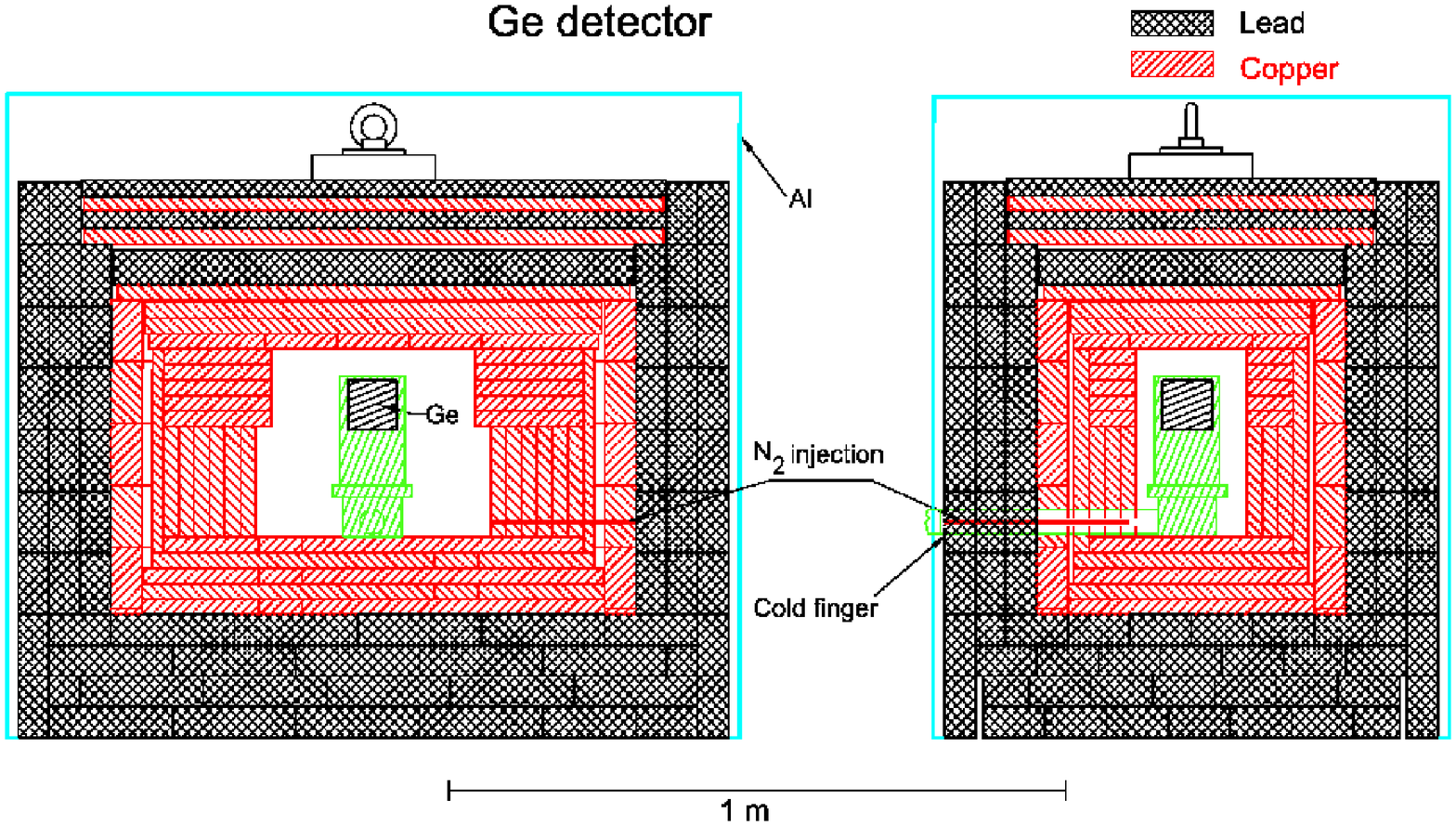}
\caption{ The germanium detector in its copper and lead shielding. The aluminum radon containment box is also shown.
}
\label{fig:gegeom}
\end{figure}

A few non-critical materials were counted in above ground detectors but most were counted underground in the Vue-des-Alpes underground laboratory. The overburden is 230~m of rock (600~m.w.e.), so that the nucleonic component of the cosmic rays is completely eliminated. The muon flux is attenuated by a factor of 1000. The detector itself is a p-type coaxial germanium detector with a useful volume of 400~ml. The geometry is indicated in Figure~\ref{fig:gegeom}.  The germanium crystal is housed in a cryostat made from highly purified Pechiney aluminum. 
All materials entering in the construction of the detector were themselves selected for low activity. The energy resolution is 2.2~keV at 2~MeV. The detector is protected against local activities by a  shielding composed of 15~cm of OFHC copper and 20~cm of lead. The shielding is contained in an air-tight aluminum box which is slightly overpressurized with boil-off nitrogen  from the detector's liquid-nitrogen dewar,  thus preventing radioactive radon gas from entering the detector volume.  As shown in Fig.~\ref{fig:gegeom}, a small volume is free around the sensitive part of the detector in order to position samples to have the best possible solid angle. Small samples were placed on top of the cryostat whereas larger ones were arranged on top and around the cylindrical part of the cryostat. This arrangement has the extra advantage that it reduces self absorption of gammas in the sample. 

\begin{figure}[htb]\centering
\includegraphics[width=\figwidth]{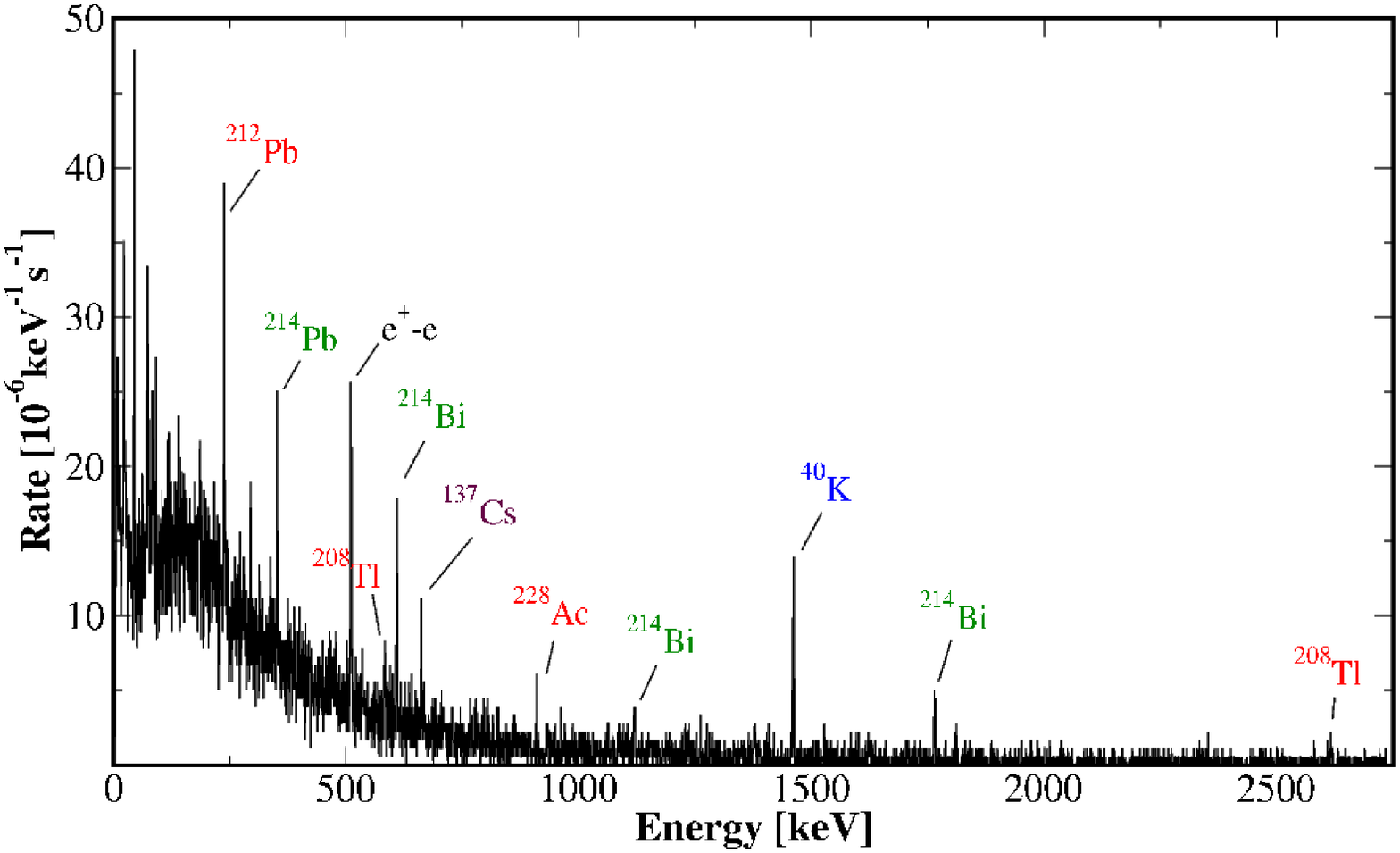}
\caption{Underground germanium background spectrum, measured with no sample for 672 h. Gamma lines from natural activities, namely the \Th{232}  (\Tl{208}, \Ac{228}, \Pb{212}) and \U{238} (\Bi{214}, \Pb{214}) chains, \K{40}, as well as the \Cs{137} line from an artificial activity, are indicated. }
\label{fig:gebkg}
\end{figure}

Prior to insertion, samples were ultrasonically cleaned in an alcohol bath. Data taking was started one day after closing the shielding to ensure that all radon gas was flushed out. Normally data were accumulated for a period of one week for each sample.  Detector backgrounds limit the effectiveness of longer accumulation times. In critical cases, (measurements of lead and copper) data were accumulated for periods as long as a month.  The contribution of a sample is obtained by subtracting the background spectrum taken without any sample (Figure~\ref{fig:gebkg}). The background was measured at regular intervals although it was found to be very stable. 

For each sample, the geometry was entered in a simplified way into a GEANT3 based Monte Carlo simulation which contained also the detector configuration. The acceptance as a function of energy of the full energy gamma peaks was computed and used to translate the observed intensity of a transition, or the upper limit on it, into a specific activity. The computed acceptance was cross checked by exposing the detector to calibrated gamma sources.  The sensitivity depends on the sample mass and configuration, which affect the solid angle and the self absorption.
For transitions with several gamma emissions contributing in parallel, or by cascade, the peak intensities were combined.  The best sensitivity was achieved with copper samples with masses of several kilograms~(see Table~\ref{TheTable} entry \idx{2}).
The quoted errors 
include statistical uncertainties as well a systematic error, dominated by the acceptance, which is estimated to be of order 10 \%.

\section{Neutron Activation Analysis}
\label{sec:NAA}
We have used  the MIT Reactor (MITR) Lab to neutron activate many material samples.  After activation and shipping, gamma emissions from the unstable activation products were observed at an EXO lab using the same germanium detector as that used for the above ground direct gamma counting mentioned in Sec~\ref{sec:gamma}.  By observing gamma energies, intensities, and decay half-lives it is possible to identify and measure the activities of many of the neutron activation products.  In practice, good spectral analysis in the presence of a multitude of gamma lines is not trivial and can have a significant impact on sensitivity and accuracy.

\begin{figure}[htb]\centering
\includegraphics[width=\figwidth]{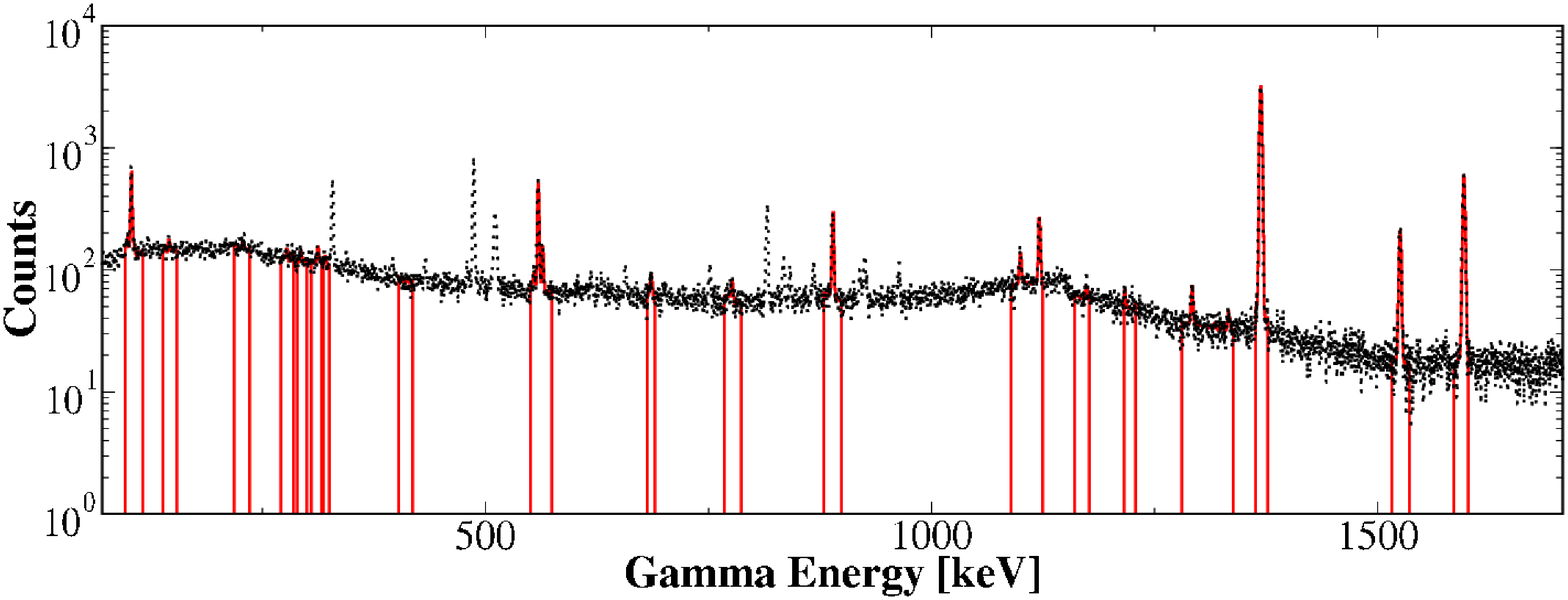}
\caption{Partial neutron-activated fly-ash gamma spectrum (dotted line) with global fit (solid line). The parameterization for this fit described backgrounds for 23 regions of interest and 62 peaks with rates linked to 31 activity parameters.}
\label{fig:flyash}
\end{figure}

\begin{figure}[htb]\centering
\includegraphics[width=\figwidth]{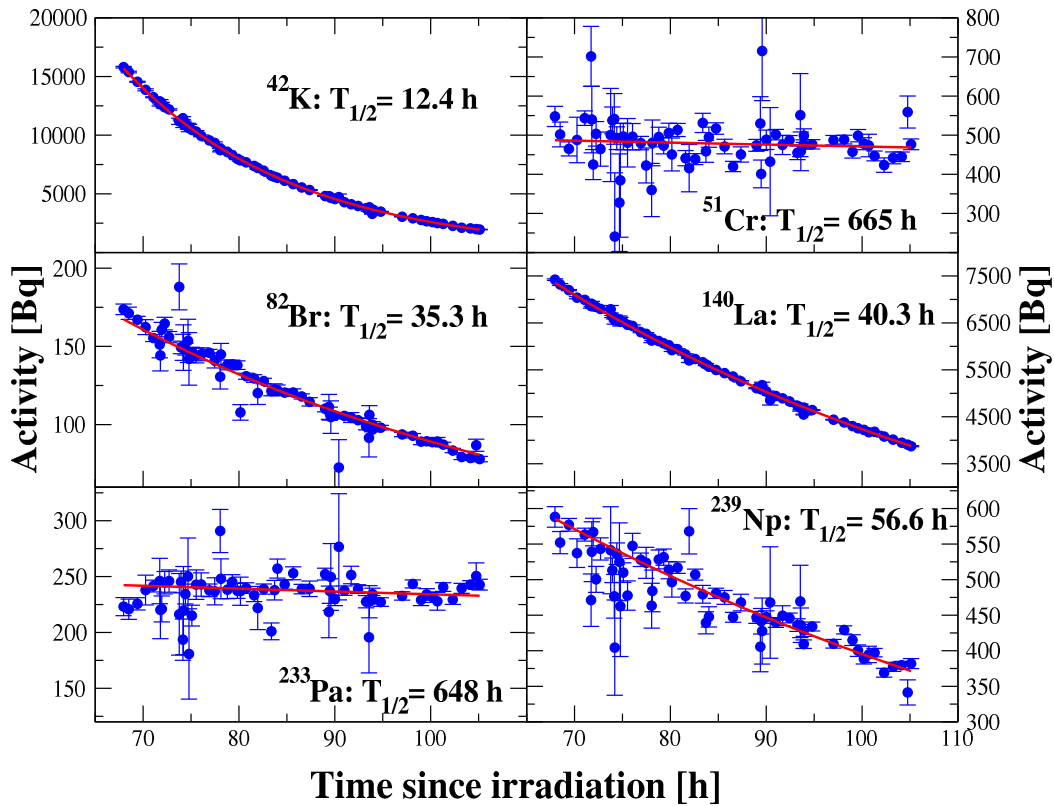}
\caption{Time development of activities in a neutron-activated fly-ash calibration sample as determined from a sequence of spectral fits (See Figure~\ref{fig:flyash}).  The points are the data;  the curves are  fit to exponential decays with the half-lives fixed to the values shown.}
\label{fig:42K}
\end{figure}

In order to maximize use of data (and thus sensitivity), as well as to simplify the analysis, a spectral fit was constructed using the most global set of experimental parameters as was practical.    Multiple energy regions of interest were defined, each having three background parameters.   Two parameters for a linear energy-calibration and two global energy-dependent peak-width parameters were used in the fit as well as one activity parameter for each isotope.   Care was taken to convert energy regions of interest to channel regions in a way which would not change the data set while minimizing the fit, which included the calibration parameters.  Peaks were defined with fixed energies, and each rate was associated with an isotopic activity via the product of the branching ratios and the detector efficiency at the relevant energy.    By parameterizing activities instead of peak areas, all peaks associated with a single isotope could be used to simultaneously constrain an activity.  The inclusion of strong, known peaks in the fits resulted in automatic energy and resolution calibration.  Interleaved calibration data could also be used to introduce constraint terms for particularly in-active samples after long delay times.  By allowing the energy calibration to float in a constrained manner while keeping peak energies fixed, subtleties of explicitly constraining peak centroids were avoided, and strong correlations in peak-position uncertainties were retained. The fit procedure was automated to produce results for data files taken in temporal sequence.  This produced time-dependent decay curves for each activity which could be fit to determine initial activities and half lives.  A calibration gamma spectrum taken from the interactive software interface is shown in Fig.~\ref{fig:flyash}.   The time series of different activities shown in Fig.~\ref{fig:42K} were found by combining the data from this and other spectra collected in sequence.


Elemental concentrations were inferred using tabulated neutron capture cross sections folded with a standard reactor neutron spectrum.  We assume that the relative isotopic abundances within the samples are given by the natural terrestrial abundance ratios.  The reactor neutron flux was periodically calibrated by activating samples of NIST coal fly-ash Standard Reference Material 1633b, which has certified known concentrations of several NAA-detectable elements including K, Th, and U.  Repeated calibrations showed a linear correlation between reactor power and thermal neutron flux.  For activation runs without a fly-ash calibration sample, this correlation was used together with the stated reactor power at the time of activation.  The fractional flux of epi-thermal neutrons was found to have only little dependence on the reactor power.  Due to the large resonance integral in the neutron capture cross section of \U{238}, this nuclide is quite sensitive to this detail.  
Results listed in Table~\ref{TheTable} include a statistical uncertainty and  a systematic uncertainty, which in most cases is dominated by a 10\% error associated with flux variations and with solid angle variations during germanium counting.  

After demonstrating good reproducibility over the course of several years, calibrations were recently performed only about once per year.  However,  the latest calibration, performed in September of 2006 showed that the ratio of total flux to epi-thermal flux was about a factor of 3 lower than found in the previous calibration, performed in March of 2005.  Data taken in the interval between these two calibrations are thus subject to significantly higher systematic uncertainties.  The affected measurements are indicated in Table~\ref{TheTable}.  The quoted results use the original 2005 calibration and do not include this added flux uncertainty.  Using the 2006 calibration, results are found to be higher by factors of approximately 1.1,  1.3, and 2.2 for K, Th and, U respectively.

NAA results using the same facility and essentially the same techniques have been reported in the past~\cite{KSCINT} in more detail.   In Ref.~\cite{KSCINT}, NAA measurement limits on the order of  $10^{-14}~g/g$ to $10^{-15}~g/g$ of Th and U were achieved for KamLAND liquid scintillator, demonstrating the potential power of the technique.  However, achieving these sensitivities required a specialized measurement campaign customized to the material of interest.  Specifically the preconcentration and chemical separation techniques used in Ref.~\cite{KSCINT} were not employed, with one exception described below, for measurements in the present work.  When surveying a large number of samples, such specialized techniques are not practical. 
Many systematic details of the technique including sample mass, containment vessel choice, and irradiation time, were varied to account for properties of individual materials and our specific analysis needs and goals.

\section{Heat Transfer Fluid}
\label{sec:HFE}

3M Novec HFE-7000, 1-methoxy\-hepta\-fluoro\-propane, serves as a heat transfer fluid and as passive shielding in EXO-200.  Because of its large mass and close proximity to the active detector volume, the allowable tolerances for U and Th contamination in this material are particularly stringent.   An extended effort was made to improve analysis sensitivities for this material.  Initially a standard neutron activation analysis was performed producing limits for  K, Th, and U concentrations which were among the lowest of any of the materials measured (see Table~\ref{TheTable} entries~~\idx{78}--\idx{81}).  We also found that HFE-7000 has very low levels of other foreign elements, making it one of the most pure materials that we have studied.  

In order to make further improvements we  preconcentrated the HFE-7000 by evaporation,  reducing its mass by a factor of 1000.  We attempted to quantify the Th and U retention of the evaporation process by mixing organo-metallic standards~\cite{KSCINT} with the HFE-7000.  These studies were unsuccessful, but indicated a very low solubility of actinides in HFE-7000.  For indirect verification of metal retention, $\rm ^{220}Rn$ gas was bubbled through an HFE-7000 mixture.  This  loaded the HFE-7000 with $\rm ^{212}Pb$ which was  detectable in a gamma detector before and after evaporation.  The concentrate was transferred from the evaporation vessels to a separate test tube and further evaporated to produce a dry residue. Approximately 20\% of the $\rm ^{212}Pb$ was retained in the concentrate residue.
With large systematic uncertainties, the amount of $\rm ^{212}Pb$  on the surfaces of the evaporation vessels was found to be consistent with the remaining 80\% of $\rm ^{212}Pb$ initially introduced.  With the assumption that this 20\% evaporation retention also applies for Th and U, we derived strict limits for their concentrations in HFE-7000 as shown in Table~\ref{TheTable} entry~\idx{82}.

\section{Shielding Lead}
\label{sec:Lead}

To find a suitable shielding lead we needed to select, as usual, for low concentrations of \KTHU, but also for low levels of \Pb{210}.  \Pb{210} and its shorter lived decay product \Po{210} are primarily beta and alpha emitters respectively.  However, because of their potentially high activities, long ranged Bremsstrahlung radiation as well as a very weak gamma line at 803~keV can potentially penetrate shielding at rates high enough to impact EXO-200 backgrounds in the energy range of the \xe two-neutrino decay mode.

To address this additional concern we used a $1700~mm^2$ ion-implanted silicon detector to observe the 5304~keV alpha particles from the decay of \Po{210} (the decay product of \Pb{210}) in several different supplies of lead. The detector efficiency was calibrated with Doe Run lead batches previously analyzed by Physikalisch-Technische Bundesanstalt (PTB) in Germany.  Background measurements were taken by replacing the lead samples with a silicon wafer. 

\begin{figure}[htb]\centering
\includegraphics[width=\figwidth]{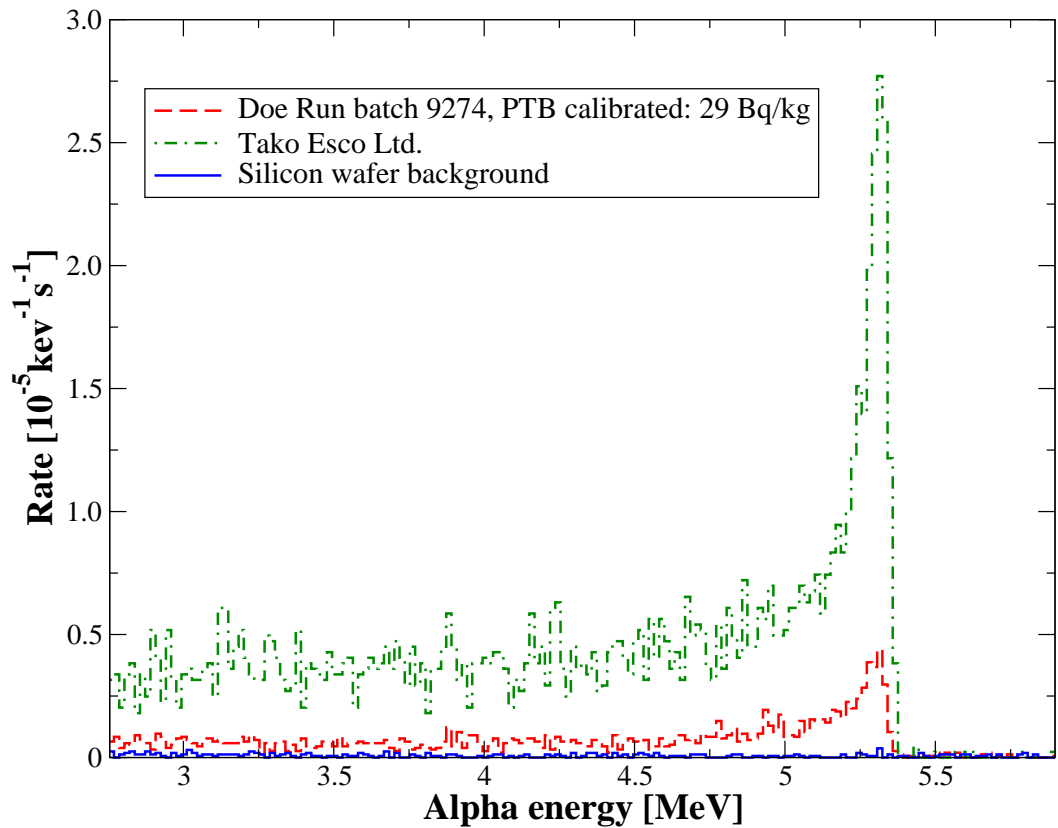}
\caption{Spectra from alpha counter including Doe Run batch 9274, which was a calibration batch measured by PTB ($\rm 28.9\pm2.9~Bq/kg$), as well the spectrum from lead supplied by Tako Esco Ltd., and a background spectrum.  The spectra are binned in groups of ten channels.}
\label{fig:lead}
\end{figure}

The EXO-200 shielding lead was selected, acquired, smelted, cleaned, and machined at JL Goslar in Germany.  Many batches of Doe Run lead were smelted together into three homogenized batches.  We analyzed the \Po{210} activities of samples from all candidate batches (about 20) of Doe Run lead purchased and from samples of all three homogenized batches.  We found that the \Po{210} activities of most candidate batches varied only within the roughly 20\% measurement uncertainty. A few batches varied by as much as 50\%, and ultimately 3 batches were thus rejected before homogenization.
In table~\ref{tab:Lead} we report the analysis results of levels of \Pb{210} in the EXO-200 lead as well as other lead varieties studied.  Alpha spectra from lead samples are shown in Figure~\ref{fig:lead} along with a background spectrum.
\scriptsize
\input{table2}
\label{tab:Lead}
\end{table}  
\normalsize

All candidate Doe Run batches and all homogenized batches were tested for \KTHU levels using GD-MS analysis. Selected batches, including all of the homogenized batches, were further tested using ICP-MS, which provided higher sensitivity to the 
uranium and thorium content. All three homogenized batches were found to have levels of K less than $\rm 7 \cdot  10^{-9}~g/g$  and both Th and U levels below  $\rm < 1\cdot 10^{-12}~g/g$. However, a few of the original candidate Doe Run batches did show levels higher than these limits.  Some of these were rejected for use in forming the EXO lead.

\section{Summary}
\label{sec:Conclusions}

Previous attempts to reduce intrinsic radioactive backgrounds in detector systems have generally focused on studies of a few common primary construction materials such as copper and lead.  In order to facilitate development of complex low-background detector systems, clean materials must be available for producing a wide variety of parts.  We have performed a systematic study of trace radioactive impurities in a large variety of parts and materials, focusing on those required to construct the EXO-200 cryogenic TPC detector system (See Table~\ref{TheTable}).  As well as finding suitable raw materials and commercially available parts, we have also shown quantitatively that efforts to improve cleanliness of the production and handling of parts can result in measurable improvements in radiopurity.  We hope that these studies will help to facilitate the advancement and development of the next generation of low-background counting facilities, both by helping to reduce background levels below what has previously been achieved, and by allowing the construction of more complex purpose-built detection systems.

\ack
We thank Dirk Arnold at PTB for contributed data, Kenji Kingsford at Saint Gobain for extensive help with Teflon selection and finishing, as well as Niel Washburn, Richard Van Ryper, and Sharon Libert at DuPont for consultations and generous materials donations.  We thank Edward Lau, Susan Tucker and Judith Maro at MITR as well as David Woisard, APT, API, and Vaga Industries for their  accommodating cooperation. This work was supported in part by the US Department of Energy under contract number DE-FG02-01ER4166.

\scriptsize
\clearpage
\onecolumn
\begin{landscape}
\setlength{\LTcapwidth}{8.5in}
\begin{longtable}[t!]{p{.2in}p{4.0in}p{.6in}p{1.1in}p{1.1in}p{1.1in}}
\caption{Measurement results for \KTHU concentrations in a variety of materials.  Manufacturer production lot numbers or arbitrary identifiers are indicated for materials where multiple lots were studied.  Uncertainties are quoted at 68\% C.L. and limits are 95\% C.L.   Results which are less than than 3-$\sigma$ above zero (not including systematic scaling uncertainties) are reported as upper limits.  GD-MS measurements have a factor of two uncertainty.   In the ``method'' column, ``A.G. Ge'' refers to above ground germanium counting.  Measurements with methods of ``Balazs Analytical Services'' or  ``Shiva Inc.'' were performed by the commercial services of the respective companies.  Entries~\protect\idx{178} and \protect\idx{226} list data taken from Refs.~\cite{darnoldprivate} and~\cite{SNO} respectively as indicated.  Where available, germanium counting results for $\rm ^{60}Co$ and $\rm ^{137}Cs$ activities are given within the sample descriptions. 
\newline$\rm ^*$ Indicated NAA results may be affected by a neutron flux calibration discrepancy described in Sec.~\ref{sec:NAA}.  The tabulated results do not include systematic uncertainties arising from this discrepancy.}
\label{TheTable}
\\\input{table3}
\end{longtable} 
\setlength{\LTcapwidth}{4in}
\end{landscape}
\normalsize
\clearpage




\bibliographystyle{elsart-num}
\bibliography{EXO_purity}






\end{document}

%% file: linemap.tex
\newcommand{\idx}[1]{%
\ife{#1}{1}	{1}%
\ife{#1}{2}	{2}%
\ife{#1}{3}	{3}%
\ife{#1}{4}	{4}%
\ife{#1}{5}	{5}%
\ife{#1}{227}	{6}%
\ife{#1}{6}	{7}%
\ife{#1}{7}	{8}%
\ife{#1}{}	{}%
\ife{#1}{}	{}%
\ife{#1}{156}	{9}%
\ife{#1}{157}	{10}%
\ife{#1}{158}	{11}%
\ife{#1}{159}	{12}%
\ife{#1}{160}	{13}%
\ife{#1}{161}	{14}%
\ife{#1}{162}	{15}%
\ife{#1}{163}	{16}%
\ife{#1}{164}	{17}%
\ife{#1}{165}	{18}%
\ife{#1}{166}	{19}%
\ife{#1}{167}	{20}%
\ife{#1}{168}	{21}%
\ife{#1}{169}	{22}%
\ife{#1}{170}	{23}%
\ife{#1}{171}	{24}%
\ife{#1}{172}	{25}%
\ife{#1}{173}	{26}%
\ife{#1}{174}	{27}%
\ife{#1}{175}	{28}%
\ife{#1}{176}	{29}%
\ife{#1}{177}	{30}%
\ife{#1}{178}	{31}%
\ife{#1}{179}	{32}%
\ife{#1}{180}	{33}%
\ife{#1}{181}	{34}%
\ife{#1}{182}	{35}%
\ife{#1}{183}	{36}%
\ife{#1}{}	{}%
\ife{#1}{}	{}%
\ife{#1}{184}	{37}%
\ife{#1}{226}	{38}%
\ife{#1}{186}	{39}%
\ife{#1}{229}	{40}%
\ife{#1}{185}	{41}%
\ife{#1}{187}	{42}%
\ife{#1}{190}	{43}%
\ife{#1}{191}	{44}%
\ife{#1}{201}	{45}%
\ife{#1}{230}	{46}%
\ife{#1}{199}	{47}%
\ife{#1}{200}	{48}%
\ife{#1}{198}	{49}%
\ife{#1}{194}	{50}%
\ife{#1}{192}	{51}%
\ife{#1}{193}	{52}%
\ife{#1}{195}	{53}%
\ife{#1}{197}	{54}%
\ife{#1}{196}	{55}%
\ife{#1}{204}	{56}%
\ife{#1}{202}	{57}%
\ife{#1}{203}	{58}%
\ife{#1}{205}	{59}%
\ife{#1}{206}	{60}%
\ife{#1}{}	{}%
\ife{#1}{}	{}%
\ife{#1}{207}	{61}%
\ife{#1}{208}	{62}%
\ife{#1}{209}	{63}%
\ife{#1}{210}	{64}%
\ife{#1}{211}	{65}%
\ife{#1}{212}	{66}%
\ife{#1}{213}	{67}%
\ife{#1}{}	{}%
\ife{#1}{}	{}%
\ife{#1}{214}	{68}%
\ife{#1}{215}	{69}%
\ife{#1}{216}	{70}%
\ife{#1}{217}	{71}%
\ife{#1}{218}	{72}%
\ife{#1}{219}	{73}%
\ife{#1}{220}	{74}%
\ife{#1}{221}	{75}%
\ife{#1}{222}	{76}%
\ife{#1}{}	{}%
\ife{#1}{}	{}%
\ife{#1}{14}	{77}%
\ife{#1}{225}	{78}%
\ife{#1}{15}	{79}%
\ife{#1}{16}	{80}%
\ife{#1}{17}	{81}%
\ife{#1}{19}	{82}%
\ife{#1}{18}	{83}%
\ife{#1}{21}	{84}%
\ife{#1}{20}	{85}%
\ife{#1}{23}	{86}%
\ife{#1}{22}	{87}%
\ife{#1}{25}	{88}%
\ife{#1}{24}	{89}%
\ife{#1}{26}	{90}%
\ife{#1}{27}	{91}%
\ife{#1}{28}	{92}%
\ife{#1}{29}	{93}%
\ife{#1}{30}	{94}%
\ife{#1}{31}	{95}%
\ife{#1}{32}	{96}%
\ife{#1}{33}	{97}%
\ife{#1}{34}	{98}%
\ife{#1}{35}	{99}%
\ife{#1}{36}	{100}%
\ife{#1}{}	{}%
\ife{#1}{}	{}%
\ife{#1}{37}	{101}%
\ife{#1}{38}	{102}%
\ife{#1}{39}	{103}%
\ife{#1}{40}	{104}%
\ife{#1}{41}	{105}%
\ife{#1}{42}	{106}%
\ife{#1}{43}	{107}%
\ife{#1}{44}	{108}%
\ife{#1}{45}	{109}%
\ife{#1}{46}	{110}%
\ife{#1}{47}	{111}%
\ife{#1}{48}	{112}%
\ife{#1}{}	{}%
\ife{#1}{}	{}%
\ife{#1}{51}	{113}%
\ife{#1}{52}	{114}%
\ife{#1}{53}	{115}%
\ife{#1}{54}	{116}%
\ife{#1}{55}	{117}%
\ife{#1}{56}	{118}%
\ife{#1}{57}	{119}%
\ife{#1}{58}	{120}%
\ife{#1}{67}	{121}%
\ife{#1}{68}	{122}%
\ife{#1}{69}	{123}%
\ife{#1}{59}	{124}%
\ife{#1}{60}	{125}%
\ife{#1}{61}	{126}%
\ife{#1}{62}	{127}%
\ife{#1}{64}	{128}%
\ife{#1}{65}	{129}%
\ife{#1}{66}	{130}%
\ife{#1}{63}	{131}%
\ife{#1}{70}	{132}%
\ife{#1}{71}	{133}%
\ife{#1}{72}	{134}%
\ife{#1}{}	{}%
\ife{#1}{}	{}%
\ife{#1}{78}	{135}%
\ife{#1}{79}	{136}%
\ife{#1}{80}	{137}%
\ife{#1}{81}	{138}%
\ife{#1}{82}	{139}%
\ife{#1}{83}	{140}%
\ife{#1}{84}	{141}%
\ife{#1}{85}	{142}%
\ife{#1}{86}	{143}%
\ife{#1}{87}	{144}%
\ife{#1}{}	{}%
\ife{#1}{}	{}%
\ife{#1}{8}	{145}%
\ife{#1}{74}	{146}%
\ife{#1}{75}	{147}%
\ife{#1}{113}	{148}%
\ife{#1}{114}	{149}%
\ife{#1}{115}	{150}%
\ife{#1}{224}	{151}%
\ife{#1}{223}	{152}%
\ife{#1}{116}	{153}%
\ife{#1}{118}	{154}%
\ife{#1}{9}	{155}%
\ife{#1}{10}	{156}%
\ife{#1}{11}	{157}%
\ife{#1}{12}	{158}%
\ife{#1}{13}	{159}%
\ife{#1}{49}	{160}%
\ife{#1}{50}	{161}%
\ife{#1}{76}	{162}%
\ife{#1}{77}	{163}%
\ife{#1}{135}	{164}%
\ife{#1}{136}	{165}%
\ife{#1}{137}	{166}%
\ife{#1}{138}	{167}%
\ife{#1}{88}	{168}%
\ife{#1}{89}	{169}%
\ife{#1}{90}	{170}%
\ife{#1}{142}	{171}%
\ife{#1}{145}	{172}%
\ife{#1}{144}	{173}%
\ife{#1}{91}	{174}%
\ife{#1}{92}	{175}%
\ife{#1}{93}	{176}%
\ife{#1}{105}	{177}%
\ife{#1}{106}	{178}%
\ife{#1}{107}	{179}%
\ife{#1}{108}	{180}%
\ife{#1}{139}	{181}%
\ife{#1}{120}	{182}%
\ife{#1}{119}	{183}%
\ife{#1}{98}	{184}%
\ife{#1}{99}	{185}%
\ife{#1}{100}	{186}%
\ife{#1}{101}	{187}%
\ife{#1}{96}	{188}%
\ife{#1}{228}	{189}%
\ife{#1}{103}	{190}%
\ife{#1}{104}	{191}%
\ife{#1}{102}	{192}%
\ife{#1}{94}	{193}%
\ife{#1}{97}	{194}%
\ife{#1}{143}	{195}%
\ife{#1}{110}	{196}%
\ife{#1}{111}	{197}%
\ife{#1}{112}	{198}%
\ife{#1}{121}	{199}%
\ife{#1}{122}	{200}%
\ife{#1}{123}	{201}%
\ife{#1}{124}	{202}%
\ife{#1}{125}	{203}%
\ife{#1}{126}	{204}%
\ife{#1}{127}	{205}%
\ife{#1}{128}	{206}%
\ife{#1}{129}	{207}%
\ife{#1}{109}	{208}%
\ife{#1}{73}	{209}%
\ife{#1}{131}	{210}%
\ife{#1}{132}	{211}%
\ife{#1}{130}	{212}%
\ife{#1}{133}	{213}%
\ife{#1}{134}	{214}%
\ife{#1}{140}	{215}%
\ife{#1}{146}	{216}%
\ife{#1}{147}	{217}%
\ife{#1}{148}	{218}%
\ife{#1}{149}	{219}%
\ife{#1}{150}	{220}%
\ife{#1}{151}	{221}%
\ife{#1}{152}	{222}%
\ife{#1}{153}	{223}%
\ife{#1}{154}	{224}%
\ife{#1}{155}	{225}%
\ife{#1}{}	{}%
\ife{#1}{230}	{225}%
\ife{#1}{}	{}%
\ife{#1}{}	{}%
\ife{#1}{}	{}%
\ife{#1}{}	{}%
\ife{#1}{}	{}%
\ife{#1}{}	{}%
\ife{#1}{}	{}%
\ife{#1}{}	{}%
\ife{#1}{}	{}%
\ife{#1}{}	{}%
\ife{#1}{}	{}%
\ife{#1}{}	{}%
\ife{#1}{}	{}%
\ife{#1}{}	{}%
\ife{#1}{}	{}%
\ife{#1}{}	{}%
\ife{#1}{}	{}%
\ife{#1}{}	{}%
}

%% file: table2.tex
 \\
\begin{table}
\begin{tabular}{lc}
{\bf Lead Source } & {\bf Activity $[Bq/kg]$ } \\
\hline
JL Goslar, Doe Run, EXO smelting lot 3-708. &   20$\pm$5  \\

JL Goslar, Doe Run, EXO smelting lot 3-707. &   17$\pm$5  \\

JL Goslar, Doe Run, EXO smelting lot 3-706. &   17$\pm$4  \\

JL Goslar, un-smelted Doe Run lot 9273. &    25$\pm$4 \\

Tako Esco Ltd. &  165$\pm$33 \\

JL Goslar, Boliden. &    18$\pm$3 \\

Plombum, VG1. & 5.9$\pm$0.9 \\

Plombum lead from Integrated Ocean Drilling Program  & 5.4$\pm$1.3 \\

JSC Industrial Corporation, lead quality C1. & 1265$\pm$183 \\
\hline
\end{tabular}
\vskip 5mm
\caption{\Po{210} (from \Pb{210} decay) activities in various lead sources as measured by alpha counting.}

%% file: table3.tex
  {\bf \#} & {\bf Material } & {\bf Method} &\centering\centering {\bf K~conc. $[10^{-9} g/g]$ } &\centering\centering {\bf Th~conc. $[10^{-12} g/g]$ } &\centering\centering {\bf U~conc. $[10^{-12} g/g]$ } \tabularnewline
           &                         \multicolumn{ 5}{c}{{\bf Bulk Copper}} \tabularnewline
           &                                    \multicolumn{ 5}{c}{{\bf }} \tabularnewline
         1 & Norddeutsche Affinerie, NOSV copper made May 2002. & Shiva Inc. GD-MS &\centering\centering        0.4 &\centering\centering      $<$5  &\centering\centering      $<$5  \tabularnewline